\documentclass[aps,reprint,floatfix,superscriptaddress]{revtex4-2}
\usepackage{geometry}
\usepackage{amsmath,amsfonts,amssymb}
\usepackage{graphicx}
\usepackage{dsfont}

\usepackage{color}
\usepackage{xcolor}

\definecolor{darkblue}{rgb}{0,0,.65}
\definecolor{darkgreen}{rgb}{0.3,0.6,0.3}
\definecolor{cyan1}{rgb}{0.0, 0.6, 0.6}
\usepackage[%
  pdfstartview=FitH,%
  breaklinks=true,%
  bookmarks=true,%
  colorlinks=true,%
  anchorcolor=black,%
  citecolor=blue,
  filecolor=black,%
  menucolor=black,%
  urlcolor=darkblue,%
  linkcolor=blue,%
 ]{hyperref}

\providecommand{\tabularnewline}{\\}

\begin{document}

\newcommand{\eg}{{\emph e.g.\ }}
\newcommand{\ie}{{\emph i.e.\ }}
\newcommand{\FIXME}[1]{{\bf FIXME: #1}}

\newcommand{\HSYK}{H_{\text{SYK}}}
\newcommand{\HBSYK}{H_{\text{b-SYK}}}

\newcommand{\up}{\uparrow}
\newcommand{\down}{\downarrow}
\newcommand\ket[1]{\left|#1\right\rangle }
\newcommand\bra[1]{\left\langle #1\right|}
\newcommand\braket[2]{\left\langle #1\middle|#2\right\rangle }
\newcommand\ketbra[2]{\left|#1\vphantom{#2}\right\rangle \left\langle \vphantom{#1}#2\right|}
\newcommand\braOket[3]{\left\langle #1\middle|#2\middle|#3\right\rangle }

\newcommand{\rSYK}{\langle r_{\text{SYK}} \rangle}
\newcommand{\rbSYK}{\langle r_{\text{b-SYK}} \rangle}
\newcommand{\rEXP}{\langle r \rangle}

\newcommand{\sgn}{\rm{sgn}}
\newcommand{\Ztwo}{{\mathbb{Z}_2}}

\title{A bipartite Sachdev-Ye-Kitaev model:\\ Conformal limit and level statistics}

\author{Mikael Fremling}

\affiliation{Institute for Theoretical Physics and Center for Extreme Matter and Emergent Phenomena,
Utrecht University, Princetonplein 5, 3584 CC Utrecht, The Netherlands}

\author{Masudul Haque}

\affiliation{Institut f\"ur Theoretische Physik, Technische Universit\"at Dresden, 01062 Dresden, Germany}

\affiliation{Department of Theoretical Physics, Maynooth University, Co. Kildare, Ireland}

\author{Lars Fritz}

\affiliation{Institute for Theoretical Physics and Center for Extreme Matter and Emergent Phenomena,
Utrecht University, Princetonplein 5, 3584 CC Utrecht, The Netherlands}

\begin{abstract}
  We study a bipartite version of the Sachdev-Ye-Kitaev (SYK) model.
  We show that the model remains solvable in the limit of large-$N$ in the same sense as the original model if the ratio of both flavors is kept finite.
  The scaling dimensions of the two species can be tuned continuously as a function of the ratio.
  We also investigate the finite-size spectral properties of the model.
  We show how the level statistics differs from the original SYK model and infer an additional exchange symmetry in the bipartite model.
\end{abstract}

\maketitle

\section{Introduction}

The Sachdev-Ye-Kitaev (SYK) model~\cite{Sachdev1993, Sachdev2015, Kitaev2015,
  Maldacena_Stanford_PRD2016, MaldacenaShenkerStanford_JHEP2016} describes a system with many
degrees of freedom with random all-to-all ($q$-body) interactions.  The original model of Sachdev
and Ye consists of pairwise coupled SU($M$) spins \cite{Sachdev1993}.  The more recent version
proposed by Kitaev \cite{Kitaev2015} has $N_{\chi}\gg1$ Majorana sites.  The $q=4$ version has the
Hamiltonian
\begin{eqnarray}\label{eq:SYK}
\HSYK=\frac{1}{4!}\sum_{i,j,l,m}J_{ijlm}\gamma_i \gamma_j \gamma_l \gamma_m\;.
\end{eqnarray}
with $N_{\chi}$ localized Majorana fermions $\gamma_i$ with $i=1,...,N_{\chi}$.
The term SYK is also used to refer to complex-fermion versions of this model and models with $q$-body interactions, with $q$ taking values other than $4$.
In this work, we will restrict to the Majorana version \eqref{eq:SYK} with four-body interactions.
The Majorana degrees of freedom have no kinetic energy in this setup; in fact, since the interactions are all-to-all, the system has zero spatial dimensions.
The interactions are usually taken to be Gaussian with mean $\langle J_{ijlm} \rangle=0$ and variance
\[
\langle J_{ijlm} J_{i^{\prime}j^{\prime}l^{\prime}m^{\prime}}\rangle=\frac{6J^2}{N_{\chi}^3}\delta_{i,i^{\prime}}\delta_{j,j^{\prime}}\delta_{l,l^{\prime}}\delta_{m,m^{\prime}}.
\] 

The SYK model has been studied intensely in the last few years, and has a number of fascinating properties.
It is a strongly coupled quantum many-body system that is maximally chaotic, as evidenced by a maximal Lyapunov exponent extracted from out-of-time-ordered correlators,
and hence acts as a fast scrambler of quantum information~\cite{Maldacena_Stanford_PRD2016,
  MaldacenaShenkerStanford_JHEP2016, Moore_Stanford_Yao_PRL2021}.
It is nearly conformally invariant, and is exactly solvable in the large $N_{\chi}$ limit~\cite{Polchinski_Rosenhaus_JHEP2016,
  Maldacena_Stanford_PRD2016, Gross_Rosenhaus_JHEP2017, Kitaev_Suh_JHEP2018, Rosenhaus_JPA2019}.
It has been used to describe two dimensional gravity and black holes~\cite{Sachdev2015, Kitaev2015,
  Maldacena_Stanford_PRD2016, Jensen_PRL2016_ChaosAdSHolography,
  Cotler_Polchinski_Shenker_Stanford_Tezuka_JHEP2017, Kitaev_Suh_JHEP2018}.
The SYK model and its extensions have also been used as a mean field model for non-Fermi liquids,
and metals without quasiparticles \cite{Davison_Georges_Sachdev_PRB2017, Kamenev_PRR2020_SYKsuperconductivity,
  Altland_Bagrets_Kamenev_PRL2019, Chubukov_PRR2020_Yukawa-SYK, Esterlis_Sachdev_PRB2021,
  Tikhanovskaya_Sachdev_Tarnopolsky_PRB2021, Franz_PRB2021_spinfulSYK}.

\begin{figure}
\includegraphics[width=0.4\textwidth]{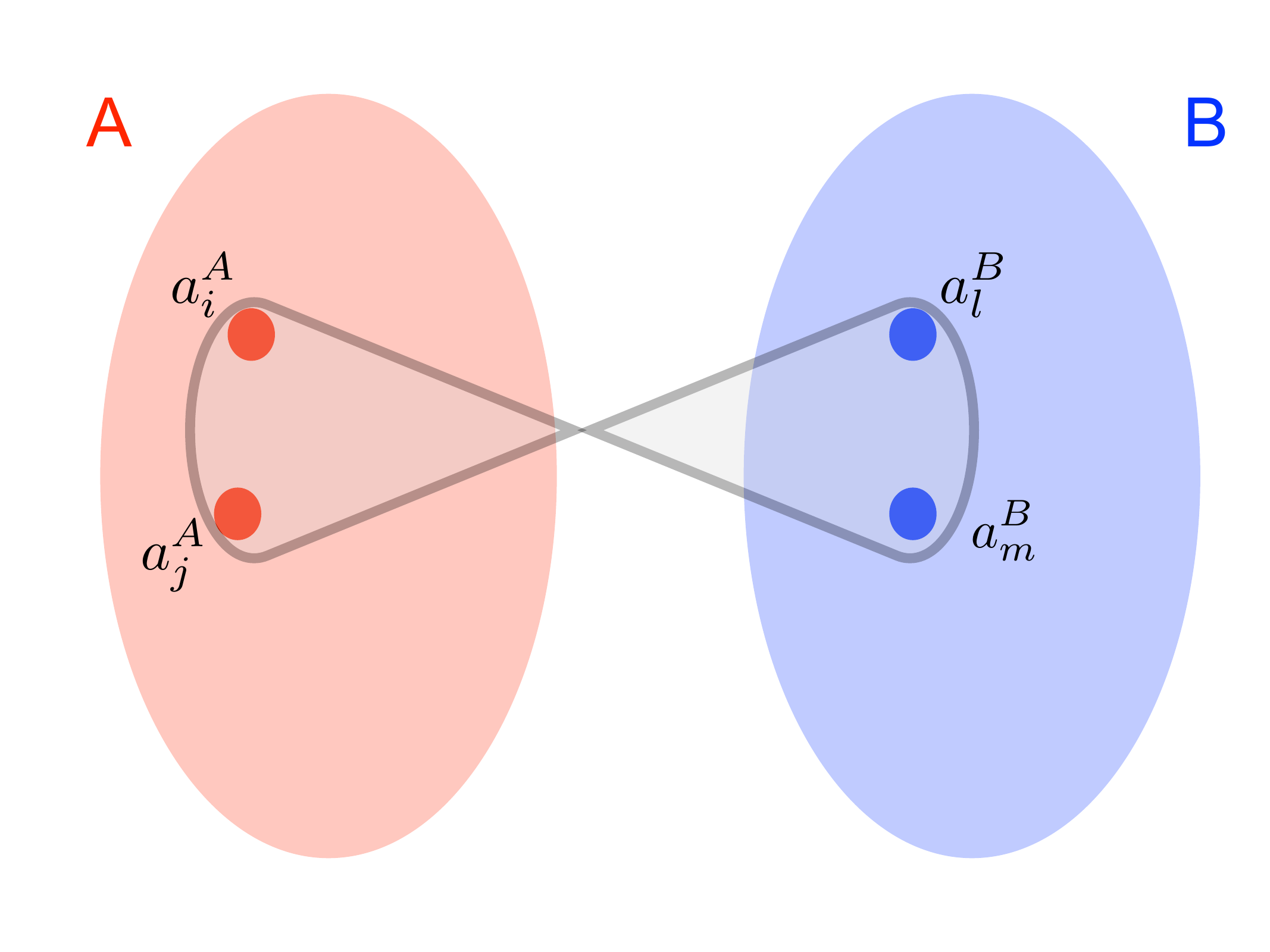}
\caption{Graphical representation of the b-SYK model. Two sets of Majorana fermions, $A$ and $B$,
do not interact within the set but strongly interact between sets. }\label{fig:bSYK}
\end{figure}

The subject of this work is a variant of the SYK model, which we henceforth refer to as the bipartite SYK (b-SYK) model.
The b-SYK was reported recently in Ref.~\cite{Fremling2021} by two of the present authors to arise as the effective
low-energy model in finite-size strained Kitaev honeycomb systems in the presence of the so-called $\Gamma$-term and moderate disorder.
The b-SYK consists of two sets of Majorana fermions, $A$ and $B$, with random 4-body interaction terms that each involve exactly two Majorana fermions from $A$ and two from $B$.
The difference with the standard SYK model is that there are no interactions within each set, only between the sets.
This is illustrated in a sketch in Fig.~\ref{fig:bSYK}.

The b-SYK model Hamiltonian is 
\begin{eqnarray}\label{eq:bSYK}
  \HBSYK = \frac{1}{4}\sum_{i,j=1}^{N_A}\sum_{l,m=1}^{N_B}
  J_{ijlm}a_{i}^Aa_{j}^Aa_{l}^Ba_{m}^B\;,
\end{eqnarray}
where $a^A$ is a Majorana fermion in set $A$ whereas $a^B$ is one in set $B$.  There are $N_A$
and $N_B$ fermions in the two sets, respectively.  The distribution of the couplings follows 
\[
\langle J_{ijlm} J_{i^{\prime}j^{\prime}l^{\prime}m^{\prime}}\rangle=\frac{J^2}{2\sqrt{N_A N_B}^3}\delta_{i,i^{\prime}}\delta_{j,j^{\prime}}\delta_{l,l^{\prime}}\delta_{m,m^{\prime}}\;.
\]

We show that the b-SYK model has an asymptotic conformal symmetry in the large-$N_{\chi}$ limit with tunable scaling dimensions
--- the scaling dimensions are a function of the relative sizes of the $A$ and $B$ sets.
By exploring the level statistics in finite-size realizations of the system,
we infer that the b-SYK system has an additional $\Ztwo$ symmetry compared to the SYK system when the two sets contain equal numbers of Majorana fermions.

In Sec.~\ref{sec:conformal} we study the large-$N_{\chi}$ limit of the
theory, deriving relations for the two-point correlator and the
scaling dimensions.  In Sec.~\ref{sec:LevelStats} we study the level
statistics of the b-SYK Hamiltonian and of the Hamiltonian
interpolating between b-SYK and SYK.  The findings of the paper and
their context are discussed in Sec.~\ref{sec:results}

\section{Some analytical properties in the conformal limit}\label{sec:conformal}

One of the features of the SYK model is that it shows conformal invariance in the infrared in the large-$N_{\chi}$ limit.
This allows for an asymptotically exact solution of its correlation functions~\cite{Maldacena_Stanford_PRD2016}.
We find that the emerging conformal symmetry of the SYK model also carries over to the b-SYK model if the large-$N_{\chi}$ limit is taken in a specific way:
the conformal symmetry exists in the limit $N_A,N_B\to \infty$ as long as the ratio $N_A/N_B=\kappa$ is kept constant.
Consequently, instead of having one scaling dimension of the Majorana fermions, like in the SYK model, the two sets of Majorana fermions, $A$ and $B$,
generally have different scaling dimensions, $\Delta_A$, and $\Delta_B$.
Their scaling dimensions depend on the parameter $\kappa$, and they can assume values between $0$ and $1/2$ while $\Delta_A+\Delta_B=1/2$.

\begin{figure}
\includegraphics[width=0.48\textwidth]{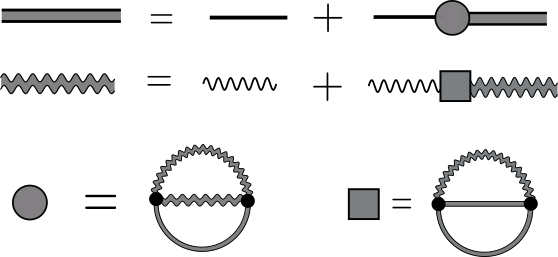}
\caption{The straight line denotes the propagator $G^{AA}$ whereas the wiggly line is $G^{BB}$.
The upper line shows the Dyson equation for $G^{AA}$ and the second line the Dyson equation for $G^{BB}$, compare Eq.~\eqref{eq:dyson_simple}.
The lowest line shows the approximation for the self-energies $\Sigma^{AA}$ and $\Sigma^{BB}$,
Eq.~\eqref{eq:self}, which becomes exact in the limit of large $N_A$ and $N_B$.}
\label{fig:feynmandiagrams}
\end{figure}

To demonstrate this, we first define the imaginary time-ordered correlation functions
\begin{equation}
\begin{split}
G^{AA}_{ij}(\tau)=\langle T_\tau \left( a^A_i(\tau) a^A_j(0)\right)\rangle\; ;\\
G^{BB}_{ij}(\tau)=\langle T_\tau \left( a^B_i(\tau) a^B_j(0)\right)\rangle\; ;\\
G^{AB}_{ij}(\tau)=\langle T_\tau \left( a^A_i(\tau) a^B_j(0)\right)\rangle\; ;\\
G^{BA}_{ij}(\tau)=\langle T_\tau \left( a^B_i(\tau) a^A_j(0)\right)\rangle\; .
\end{split}
\end{equation}
The Green function of the non-interacting problem is given by
\begin{equation}\label{eq:bareprop}
\begin{split}
G^{AA}_{0,ij}(\tau)&=\frac{1}{2}\sgn (\tau) \delta_{i,j}\; , \\
G^{BB}_{0,ij}(\tau)&=\frac{1}{2}\sgn (\tau) \delta_{i,j}\;  \\
G^{AB}_{0,ij}(\tau)= G^{BA}_{0,ij}(\tau)&=0\; . 
\end{split}
\end{equation}
meaning it is local in the index $i,j$ as well as the set label $A,B$.
It constitutes the starting point for the perturbation theory to follow.
The most general Dyson equation reads
\begin{widetext}
\begin{eqnarray}\label{eq:dyson}
  \int d \tau'  \left (\begin{array}{cc}  G^{AA-1}_{0,ij}(\tau,\tau')-\Sigma^{AA}_{ij}(\tau,\tau')
    &G^{AB-1}_{0,ij}(\tau,\tau') -\Sigma^{AB}_{ij}(\tau,\tau') \\
    G^{BA-1}_{0,ij}(\tau,\tau')-\Sigma^{BA}_{ij}(\tau,\tau')
    & G^{BB-1}_{0,ij}(\tau,\tau')-\Sigma^{BB}_{ij}(\tau,\tau')   \end{array} \right)
  \left( \begin{array}{cc} G^{AA}_{jk}(\tau',\tau'') & G^{AB}_{jk}(\tau',\tau'') \\
    G^{BA}_{jk}(\tau',\tau'') & G^{BB}_{jk}(\tau',\tau'')  \end{array} \right)=
  \delta(\tau-\tau'')\delta_{i,k}\mathds{1}\nonumber \\
\end{eqnarray}
\end{widetext}
where $\Sigma^{AA}$, $\Sigma^{AB}$, $\Sigma^{BA}$, and $\Sigma^{BB}$ are the self-energies whereas $G^{AA}$, $G^{AB}$, $G^{BA}$, and $G^{BB}$ are the Green functions.
Summation over double indices is implied.
In general, this equation is non-local in both the indices $i,j$ as well as the set labels $A,B$.
The most transparent way to determine the self-energies is based on a diagrammatic representation of perturbation theory in terms of Feynman diagrams.
To leading order in $N_A$ and $N_B$, the diagrams shown in Fig.~\ref{fig:feynmandiagrams} constitute the entire perturbative series and can be resummed exactly.
This implies that in this limit, the theory remains local in $i,j$ as well as $A,B$.
Consequently, to leading order the off-diagonal self-energies
$\Sigma^{AB}_{ii}(\tau_1,\tau_2)$ as well as the off-diagonal Green
functions $G^{AB}$ and $G^{BA}$ vanish.  
We can drop the subscripts $i,j$, as all $\Sigma$'s and
$G$'s are diagonal in these indices and independent of $i$.
Furthermore, the self-energies $\Sigma^{AA}_{ii}(\tau_1,\tau_2)$ and
$\Sigma^{BB}_{ii}(\tau_1,\tau_2)$ dominate the bare propagators,
Eq.~\eqref{eq:bareprop}, in the infrared.  

Thus, the Dyson equation reduces to 
\begin{widetext}
\begin{eqnarray}\label{eq:dyson_simple}
  \int d \tau' \left (\begin{array}{cc} \Sigma^{AA}(\tau,\tau')G^{AA}(\tau',\tau'') &0\\
    0& \Sigma^{BB}(\tau,\tau')G^{BB}(\tau',\tau'')    \end{array}
  \right)  =-\delta(\tau-\tau'')\mathds{1}\;.\nonumber \\
\end{eqnarray}
Utilizing translational invariance in time and passing over to
frequency space using a Fourier
transformation, we obtain
\begin{equation}  \label{eq:dyson_simple_freqspace}
 \Sigma^{AA}\left(\omega\right)G^{AA}\left(\omega\right)=-1; \qquad 
 \Sigma^{BB}\left(\omega\right)G^{BB}\left(\omega\right)=-1.
\end{equation}

We now consider the leading-order approximation shown in the diagrams
of Figure \ref{fig:feynmandiagrams}:  
\begin{equation}\label{eq:self}
\begin{split}
  \Sigma^{AA}(\tau)&=\frac{J^2N_AN_B^2}{\sqrt{N_A^3N_B^3}} G^{AA}(\tau)G^{BB}(\tau)G^{BB}(\tau) 
  =\frac{J^2}{\sqrt{\kappa}} G^{AA}(\tau)G^{BB}(\tau)G^{BB}(\tau)   \\
  \Sigma^{BB}(\tau)&=\frac{J^2N_A^2N_B}{\sqrt{N_A^3N_B^3}} G^{BB}(\tau)G^{AA}(\tau)G^{AA}(\tau)
  =J^2\sqrt{\kappa} G^{BB}(\tau)G^{AA}(\tau)G^{AA}(\tau)\;.
\end{split}
\end{equation}
\end{widetext}
Because of translational invariance in time, all $G$'s and $\Sigma$'s
only contain relative coordinates; hence they each have a single time
argument instead of two.

Due to the time reparametrization symmetry of the theory we expect conformal invariance.
Since we are expecting different scaling dimensions for the Majorana fermions in the two sets we introduce the scaling dimension $\Delta_A$ for Majorana fermions in set $A$,
whereas we introduce $\Delta_B$ for those in set $B$.
We then have
\begin{equation}\label{eq:GrFns_conformalansatz}
\begin{split}
  G^{AA}(\tau)&= C_A \frac{\sgn(\tau)}{|\tau|^{2\Delta_A}},  \\
  G^{BB}(\tau)&= C_B \frac{\sgn(\tau)}{|\tau|^{2\Delta_B}},
\end{split}
\end{equation}
for the full Green functions.  Here $C_A$ and $C_B$ are constants.  Inserting this conformal anzatz into
Eqs.\ \eqref{eq:self}, the self energies read
\begin{equation}\label{eq:selfenergies_from_conformalansatz}
\begin{split}
\Sigma^{AA}\left(\tau\right) &
=\frac{1}{\sqrt{\kappa}}J^2 C_A C_B^2\frac{\sgn\left(\tau\right)}{\left|\tau\right|^{2\Delta_A+4\Delta_B}}, 
\\
\Sigma^{BB}\left(\tau\right) & =\sqrt{\kappa}J^2 C_B C_A^2\frac{\sgn\left(\tau\right)}{\left|\tau\right|^{2\Delta_B+4\Delta_A}},
\end{split}
\end{equation}
where $\kappa=N_A/N_B$.

We now express Eqs.\ \eqref{eq:GrFns_conformalansatz} and
\eqref{eq:selfenergies_from_conformalansatz} in frequency space.  Using the identity
\begin{equation}
  \int_{-\infty}^{\infty} d\tau e^{\imath \omega \tau}\frac{\sgn(\tau)}{|\tau|^{\alpha}}=
  \imath \sqrt{\pi}\frac{\Gamma \left(
  1-\frac{\alpha}{2}\right)}{\Gamma \left(\frac{1}{2}+\frac{\alpha}{2}
  \right)}\sgn(\omega)
\left(\frac{\left|\omega\right|}{2}\right)^{\alpha-1} , 
\end{equation}
we obtain for the Green functions
\[
G^{AA}\left(\omega\right) =
C_A\imath\sqrt{\pi}\frac{\Gamma\left(1-\Delta_{A}\right)}{\Gamma\left(\frac{1}{2}+\Delta_{A}\right)}
\sgn\left(\omega\right)\left(\frac{\left|\omega\right|}{2}\right)^{2\Delta_{A}-1}
, 
\]
and similarly for $G^{BB}\left(\omega\right)$.
Applying the Fourier transform to Eqs.\
\eqref{eq:selfenergies_from_conformalansatz} and using the same
identity, we obtain for the self energies 
\begin{align*}
  \Sigma^{AA}\left(\omega\right)  =&\frac{J^{2}}{\sqrt{\kappa}} C_A C_B^{2}\imath\sqrt{\pi}
  \frac{\Gamma\left(1-\Delta_{A}-2\Delta_{B}\right)}{\Gamma\left(\frac{1}{2}+\Delta_{A}+2\Delta_{B}\right)}\nonumber\\
  &\quad\times\sgn\left(\omega\right)\left(\frac{\left|\omega\right|}{2}\right)^{2\Delta_{A}+4\Delta_{B}-1}
    , 
\end{align*}
and similarly for $\Sigma^{BB}$ with $\kappa\to1/\kappa$.

We can now use these expressions in our frequency-space Dyson
equation, Eq.\ \eqref{eq:dyson_simple_freqspace}.  The first equation
(for $A$) then reads  
\begin{align*}
  -1= & -\frac{J^{2}}{\sqrt{\kappa}} C_A^{2} C_B^{2}\pi
  \left(\frac{\left|\omega\right|}{2}\right)^{4\Delta_{A}+4\Delta_{B}-2}\\
  &\quad\times\frac{\Gamma\left(1-\Delta_{A}-2\Delta_{B}\right)}{\Gamma\left(\frac{1}{2}+\Delta_{A}+2\Delta_{B}\right)}\frac{\Gamma\left(1-\Delta_{A}\right)}{\Gamma\left(\frac{1}{2}+\Delta_{A}\right)}.
\end{align*}
Since the left-hand side is $\omega$-independent, we need
to remove the $\omega$ dependence
on the right-hand side.  Imposing the condition leads to the result
\begin{equation}
\Delta_{A}+\Delta_{B}=\frac{1}{2}
\end{equation}
as announced at the beginning of this section.  

\begin{figure}[t]
\includegraphics[width=0.48\textwidth]{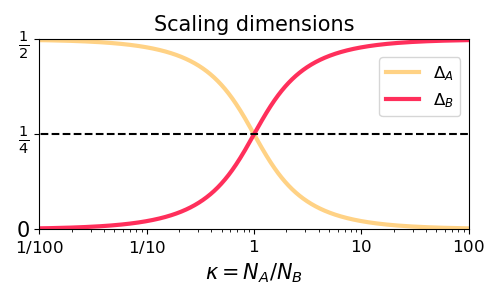}
\caption{Scaling dimensions $\Delta_A$ and $\Delta_B$ as a function of $\kappa$.
Both interpolate between $0$ and $1/2$.}\label{fig:scalingdim}
\end{figure}

Defining $\Lambda=\pi J^{2}C_A^{2}C_B^{2}$, the Dyson equations now read
\begin{equation} \label{eq:Dysoneqs_B}
\begin{split}
1= & \frac{\Lambda}{\sqrt{\kappa}}\frac{\Gamma\left(1-\Delta_{A}-2\Delta_{B}\right)}{\Gamma\left(\frac{1}{2}+\Delta_{A}+2\Delta_{B}\right)}\frac{\Gamma\left(1-\Delta_{A}\right)}{\Gamma\left(\frac{1}{2}+\Delta_{A}\right)},\\
1= & \sqrt{\kappa}\Lambda\frac{\Gamma\left(1-2\Delta_{A}-\Delta_{B}\right)}{\Gamma\left(\frac{1}{2}+2\Delta_{A}+\Delta_{B}\right)}\frac{\Gamma\left(1-\Delta_{B}\right)}{\Gamma\left(\frac{1}{2}+\Delta_{B}\right)}
.
\end{split}
\end{equation}
By eliminating $\Lambda$ and using properties of the Gamma function
(Appendix \ref{app_scaling_dims}), we can relate $\kappa$ to the
scaling dimensions ($\Delta_A = \frac{1}{2}-\Delta_B$):
\begin{equation} \label{eq_scalin_dims}
  \kappa = \frac{2 \Delta_A}{1-2\Delta_A}\left( \frac{1}{\tan \left( \pi \Delta_A\right)}\right)^2\;.
\end{equation}
This equation implicitly provides the scaling dimension $\Delta_A$
(and hence also $\Delta_B$) as a function of the ratio of sizes of the
two partitions, $\kappa=N_A/N_B$.    
For $\kappa=1$, we find $\Delta_A=\Delta_B=1/4$, as expected, just like in the standard SYK model.
For other values of $\kappa$, both scaling dimensions interpolate between $0$ and $1/2$ while always fulfilling $\Delta_A+\Delta_B=1/2$.
This behavior is presented in Fig.~\ref{fig:scalingdim} on a logarithmic scale which shows the $A$-$B$ symmetry explicitly.
Tunable scaling dimensions have also been seen in other variants of the SYK model \eg Ref.~\cite{Marcus2019,Kim2019}.

Due to the conformal invariance and the reparametrization invariance it is straightforward
to determine the finite temperature and real time correlators. At finite temperatures we find
\begin{eqnarray}
G^{AA}(\tau)&=&A \; \sgn(\tau) \left(  \frac{\pi}{\beta \sinh\left(\frac{\pi \tau}{\beta} \right)}\right)^{2\Delta_A}\;,\nonumber \\ G^{BB}(\tau)&=&B\;  \sgn(\tau) \left(  \frac{\pi}{\beta \sinh\left(\frac{\pi \tau}{\beta} \right)}\right)^{2\Delta_B}\;,
\end{eqnarray}
whereas for the retarded propagator at finite temperature we obtain
\begin{eqnarray}
G^{AA}_{\rm{ret}}(t)&=&\theta(t) A \cos \left(\pi \Delta_A \right)\left(  \frac{\pi}{\beta \sinh\left(\frac{\pi t}{\beta} \right)}\right)^{2\Delta_A} \;,\nonumber \\ G^{BB}_{\rm{ret}}(t)&=&\theta(t)B \cos \left(\pi \Delta_B \right)\left(  \frac{\pi}{\beta \sinh\left(\frac{\pi t}{\beta} \right)}\right)^{2\Delta_B}\;.\nonumber \\
\end{eqnarray}

\section{Level statistics}\label{sec:LevelStats}

In this section, we focus on the level spacing statistics of the b-SYK model.
For this purpose, we consider finite $N_A$ and $N_B$, and diagonalize the many-body SYK and b-SYK Hamiltonians.

Level statistics can help identify the existence of chaos (non-integrability) in quantum Hamiltonians,
and also to distinguish between different symmetry classes.
The interest in the SYK model is partly due to its being maximally chaotic.
Therefore, eigenvalue statistics has been a widely used diagnostic for characterizing the SYK model
\cite{GarciaGarcia_Verbaarschot_PRD2016,
  You2017, Garcia_Verbaarschot_PRD2017, Cotler_Polchinski_Shenker_Stanford_Tezuka_JHEP2017,
  Haque2019, Behrends_Bardarson_Beri_PRB2019} and its various variants \cite{Milekhin_2021, Li_Liu_Zhou_susySYK_JHEP2017, Kanazawa_Wettig_JHEP2017,
  GarciaGarcia_Tezuka_PRL2018, Katsura_Sagawa_PRD2018, Haque2019,
  GarciaGarcia_Nosaka_Rosa_Verbaarschot_PRD2019, SunYe_PRL2020, SunYeLiu_PRD2020,
  Nosaka_Numasawa_JHEP2020_massdeformedSYK, Behrends_Beri_PRD2020, Behrends_Beri_PRL2020,
  Galitski_syk2_PRL2020, Lau_Tezuka_JPhysA2021, 
  GarciaGarcia_Jia_Rosa_Verbaarschot_PRD2021, Sa_GarciaGarcia_2021_WishartSYK,
  GarciaGarcia_Sa_Verbaarschot_2021_nonhermSYK, Sun_YiXiang_Ye_Liu_PRB2021_chaotictointegrable}.
A noteworthy feature of the SYK level statistics is that it depends on the number of Majorana fermions $N_{\chi}$.
We show below that the level statistics of the b-SYK model in the $N_A=N_B$ case is systematically shifted with respect to that of the standard SYK model,
consistent with the presence of an extra $\Ztwo$ symmetry in the b-SYK system.

\subsection{Relevant ensembles}\label{sec:ensembles_r}

The universality classes of random matrices that are relevant for us include the Gaussian Orthogonal Ensemble (GOE),
the Gaussian Unitary Ensemble (GUE), and the Gaussian Symplectic Ensemble (GSE).
In Table \ref{tab:LevStat} we refer to these as O, U, and S respectively for conciseness.
Additionally, we will encounter below the level statistics obtained by merging the spectra of two GOE matrices;
we refer to this as $2\times$GOE, or for conciseness 2O in Table \ref{tab:LevStat}.

\begin{table}[t]
  \begin{tabular}{c|c|c|c|c|c|c|c|c}
    $N_{\chi}$ (mod 8) & 0 & 1 & 2 & 3 & 4 & 5 & 6 & 7\tabularnewline
    \hline 
    $\HSYK$
    & O & O & U & S & S & S & U & O\tabularnewline
    $\HBSYK$
    & 2O & 2O & O & U & U & U & O & 2O\tabularnewline
  \end{tabular}
  \caption{Level statistics of the Majorana fermion SYK model,
$\HSYK$ as compared to $\HBSYK$ in \eqref{eq:bSYK}.
    Here O=GOE, U=GUE, S=GSE and 2O = 2$\times$GOE are the different universal random matrix ensembles.
    Note how the level statistics of $\HBSYK$ traces those of $\HSYK$ but with a reduction in the symmetry classifications of one step such that S $\to$ U $\to$ O $\to$ 2O.}
  \label{tab:LevStat}
\end{table}

For characterizing the level statistics with a single number, it has become common to use the average ratio of successive level spacings~\cite{Oganesyan2007,Atas2013}.
One starts with calculating the finite size spectrum $E_n$, which are ordered from lowest to highest energy.
The set of level spacings are defined as $s_n = E_{n+1}-E_n$.  This allows to define the ratio
\begin{equation}
r_n = \frac{{\rm min}(s_n, s_{n-1})}{{\rm max}(s_n, s_{n-1})}\;.
\end{equation}

Analyzing the statistics of this quantity $r_n$ has advantages over the statistics of the bare level spacings $s_n$ themselves.
It bypasses the need to account for a varying density of states through unfolding procedures.
In addition, the average of this quantity has characteristic values for the different ensembles,
thus enabling one to distinguish symmetry classes without analyzing complete distributions.

For the Wigner-Dyson ensembles, the probability distributions of the ratio $r$ are well-approximated
by the surmise~\cite{Atas2013} $P(r) \propto (r+r^2)^{\beta}/(1+r+r^2)^{1+3\beta/2}$ up to
normalization, with $\beta=1$ for GOE, $\beta=2$ for GUE, and $\beta=4$ for GSE.
The averages of these distributions are found to be $\langle r \rangle_{\rm GOE}\approx 0.53$,
$\langle r \rangle_{\rm GUE}\approx 0.60$, and $\langle r \rangle_{\rm GSE}\approx 0.67$~\cite{Atas2013}.

Integrable (non-chaotic) Hamiltonians, which do not show level repulsion, generically have Poisson statistics,
for which the level spacing ratio has probability distribution $P(r)=2/(1+r)^2$ and mean value $\langle r \rangle=2\ln 2-1 \approx 0.39$.
An integrable system can be thought of as having a large number of conserved quantities or quantum numbers.
Therefore, adding one or a few conservation laws to a GOE system is expected to change the distribution to a form intermediate between the GOE and Poisson cases.
The $2\times$GOE spectrum can be interpreted as that obtained when a GOE system acquires a single quantum number with two possible values,
which splits the GOE spectrum into two sectors.
Thus, we expect its level spacing distribution to be intermediate between Poisson and GOE distributions.
Indeed, we find numerically, by merging the spectra of two GOE matrices, that the $2\times$GOE distribution has $\langle{r}\rangle \approx 0.425$, intermediate between the Poisson and GOE values.
Some analytic formulas for the $2\times$GOE distribution were also provided in Refs.~\onlinecite{Giraud_Vernier_Alet_2020,Fremling2022}.

As discrete symmetries are common in quantum Hamiltonians, spectra formed out of two or more independent GOE or GUE components are
the subject of longstanding interest in the quantum-chaos and random-matrix literature
\cite{Rosenzweig_Porter_PR1960, Guhr_Weidenmuller_ChemPhys1990,
  Hartmann_Weidenmuller_Guhr_ChemPhys1991, Ma_JPSJ1995_CorrHole, Alt_etal_PRE1997_correlationhole,
  Guhr_MuellerGroeling_WeidenMueller_PhysRep1998, Reichl_book2004,
  Molina_Relano_Faleiro_PhysLettB2007, Weidenmueller_Mitchell_RMP2009, Haake_book2010,
  delaCruz_Lerma_Hirsch_PRE2020_symmetries, SunYeLiu_PRD2020,
  Tekur_Santhanam_PRR2020, Giraud_Vernier_Alet_2020, Fremling2022}.
Here, we will only be concerned with the $2\times$GOE case because restricting the couplings of the
SYK Hamiltonian to obtain the b-SYK Hamiltonian effectively adds a single $\Ztwo$ symmetry.

\begin{figure}[t]
  \begin{centering}
    \includegraphics[width=0.95\linewidth]{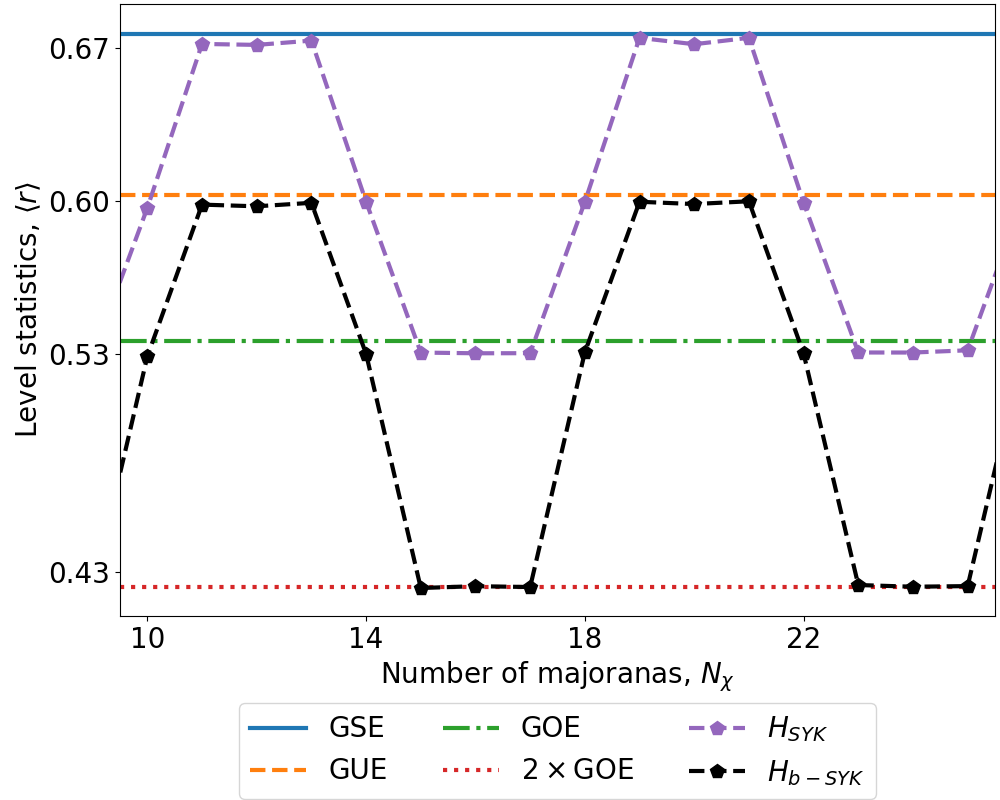}
    \par\end{centering}
    \caption{Average  level spacing ratio $\rEXP$ for $\HSYK$ and $\HBSYK$, plotted against system size $N_{\chi}$.
      For each system size, the value of  $\rEXP$ for $\HBSYK$ is  lower than that  for $\HSYK$.
      The SYK-model has an 8-fold periodicity \cite{You2017, Garcia_Verbaarschot_PRD2017,   Cotler_Polchinski_Shenker_Stanford_Tezuka_JHEP2017, Haque2019}
      related to the Bott periodicity.
      The overall  $\rEXP$ vs $N_{\chi}$ behavior for $\HBSYK$ mimics that for $\HSYK$ but with a shifted classification.
      Thus instead of the SYK sequence we get a relative shift as O$\to$2O, U$\to$O and S$\to$U.
      \label{fig:LevStat}}
\end{figure}

\subsection{Level statistics of b-SYK}\label{sec:levelstatbsyk}

In the case of the SYK model, the random matrix ensemble describing the level statistics changes with the number of Majorana fermions $N_\chi$
\cite{You2017, Garcia_Verbaarschot_PRD2017,   Cotler_Polchinski_Shenker_Stanford_Tezuka_JHEP2017, Haque2019} as listed in Table~\ref{tab:LevStat}.
This dependence is cyclic modulo $8$ in $N_\chi$ and is related to the 8-fold Bott-periodicity.
Here we will compare the level spacing statistics properties of the SYK model to that of the b-SYK model.
We will concentrate on the case $\kappa=1$, and we choose $N_\chi=2N_a=2N_b$ or $N_\chi=2N_a+1=2N_b-1$ depending on the parity of $N_\chi$.

The numerical procedure used to obtain level statistics is described
briefly in Appendix \ref{app_numerics}.

We quantify the level statistics by the average ratio $\langle{r}\rangle$, described above.
Some results are summarized in Fig.~\ref{fig:LevStat}, for both the SYK and the b-SYK Hamiltonians.
For each $N_\chi$, the averaging of the spacing ratio is performed over the spectra of many coupling realizations so that the results are sufficiently converged.
In Fig.~\ref{fig:LevStat} the horizontal lines represent the average $r$ values for different relevant ensembles, as discussed above.

We observe that the average spacing ratio of the b-SYK model is always lower than the average spacing ratio expected from the SYK model, irrespective of the size of the system.
However, it follows the same 8-fold periodicity in the total number of Majorana fermions, $N_\chi$.
Compared to the SYK sequence, we find relative shifts O$\to$2O, U$\to$O, S$\to$U, i.e., the GSE, GUE, and GOE get converted to GUE, GOE, and $2\times$GOE respectively.
The shift is also seen by comparing the two rows of Table \ref{tab:LevStat}.

\begin{figure}[t]
  \begin{centering}
    \includegraphics[width=0.95\linewidth]{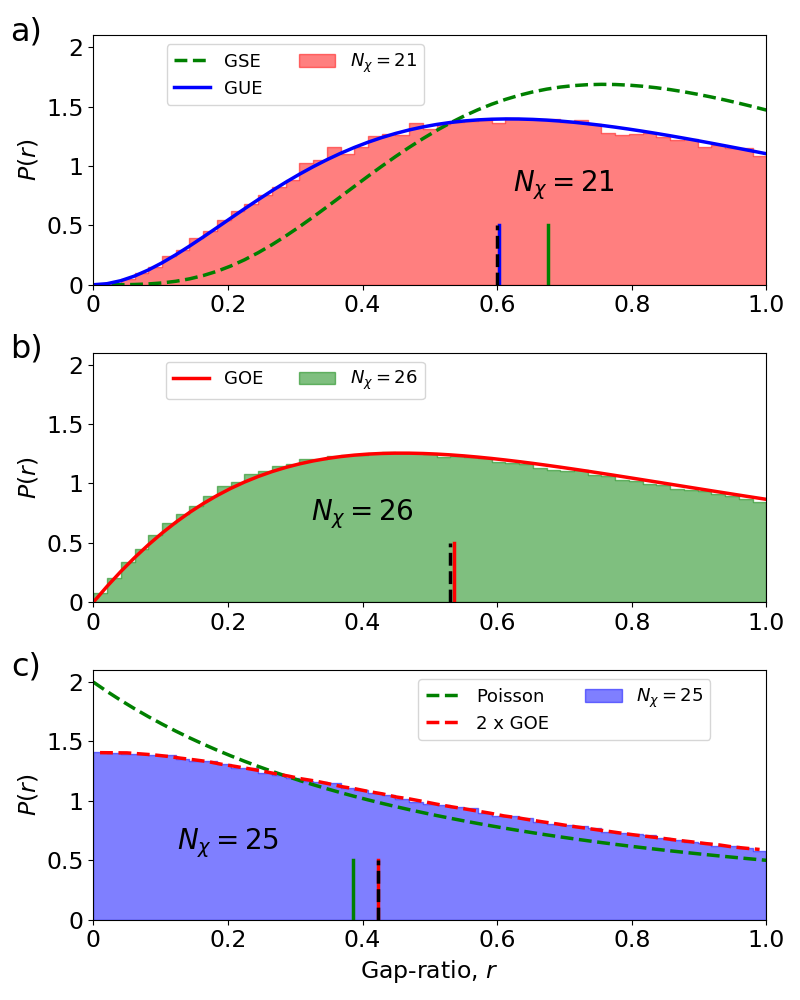}
    \par\end{centering}
  \caption{Distribution of level spacing ratios for the b-SYK Hamiltonian with $N_\chi=21,25,26$,
    corresponding to the three distinct classes listed in Table \ref{tab:LevStat}.
    The distributions for a) $N_\chi=21$ and b) $N_\chi=26$  match well the expected surmise for the GUE and GOE classes.
    The third example, c) $N_\chi=25$, closely follows the 2$\times$GOE distribution.
    The GSE and Poissonian distributions are also shown as lines.
    None of the b-SYK distributions follow the GSE class (in contrast to the SYK case) or the Poissonian distribution.
    The vertical lines are a visual representation of the mean $\left<r\right>$ for the various distributions.
    \label{fig:GapRadioDist}}
\end{figure}

Going beyond the average, in Figure \ref{fig:GapRadioDist} we show the full distributions (numerical histograms)
of the level spacing ratio, for the $\HBSYK$ Hamiltonian with $N_\chi=21,25,26$.
Clearly, the three classes follow the expected distributions for $2\times$GOE, GOE, and GUE, shown as dotted lines.
The GSE distribution is not obtained in the b-SYK system for any value of $N_{\chi}$.  

The shift in level statistics relative to SYK is clearly due to the restriction to bipartite interactions.
The results are consistent with the explanation that the bipartite structure leads to an additional $\Ztwo$ symmetry of the Hamiltonian.
A system of the GUE symmetry class, if endowed with an additional $\Ztwo$ symmetry, shows GOE level statistics~\cite{Robnik1986, Berry1986, Izrailev1990, Seligman1985}.
This effect was discussed early in the context of single-particle (billiard) systems with a magnetic field~\cite{Robnik1986, Berry1986}.
This system would naively be expected to have GUE statistics due to broken time-reversal symmetry.
However, when reflection symmetry is present, the level statistics is of the GOE class.
This phenomenon has also been observed in a many-body system~\cite{Fremling2018}.
In the present case, the anti-unitary symmetry involved is not time, but the effect is the same:
For $N_\chi=10,14,18,22\ldots$, the SYK level statistics are GUE, but the b-SYK level statistics are of GOE type.

For values of $N_\chi$ for which the level statistics if of GSE type, a corresponding effect is seen.
The additional symmetry reduces the degree of level repulsion, and one obtains GUE statistics instead, as seen in Fig.~\ref{fig:LevStat} and Table \ref{tab:LevStat}.
A GSE to GUE shift due to a parity symmetry is discussed in Section 2.7 of Ref.~\cite{Haake_book2010}.
Still, we do not know of another example in the literature involving a many-body Hamiltonian.
The b-SYK spectrum retains the Kramers degeneracy; the level repulsion is between pairs of degenerate states.

\subsection{The $\Ztwo$ symmetry}\label{sec:which_Z2}

The numerical data implies that the b-SYK Hamiltonian possesses a
$\Ztwo$ symmetry which is not present in the SYK model.

The extra symmetry arises because the b-SYK restriction removes terms
in the Hamiltonian that has an odd number of $A$ fermions, or an odd
number of $B$ fermions.  Each b-SYK term has exactly two $A$ fermions and
exactly two $B$ fermions.  Because each term is bilinear in the
$a_j^A$'s as well as in the $a_j^B$'s, if one flips the signs of all
the $A$ operators, while keeping all the $B$ operators fixed, each term
in the Hamiltonian would remain unchanged.  

Thus, the b-SYK model admits a global sign flip symmetry $a_j^A\to-a_j^A$,
$a_j^B\to+a_j^B$. 
In contrast to the b-SYK model, the SYK model is not invariant under
this transformation, as  the usual SYK model  also contains
terms of the form $a_i^Aa_j^Ba_k^Ba_l^B$ and $a_i^Aa_j^Aa_k^Aa_l^B$.  
Since these terms have an odd number of both $a^A$ and $a^B$
Majoranas, they would change sign under a sign change of  only $a^A$'s
(or a sign change of only $a^B$'s).

For even $N_A$, the sign flip operation can be expressed in terms of
the Hermitian operator
\begin{equation}
  \Gamma = \imath^{N_A/2}\prod_{i=1}^{N_A} a^A_i.
\end{equation}
Using Majorana anticommutation relations, one finds that this operator
satisfies $\Gamma a^A_j= -a^A_j\Gamma$, provided that 
$N_A$ is even.  For odd $N_A$, an additional fictitious Majorana has to be added
to the product to construct a Hermitian sign-flip operator.  

Equivalently, one could  flip the signs of the $B$ Majorana
operators, and keep the b-SYK Hamiltonian invariant.  This is not an
independent extra symmetry compared to the SYK model, as the flipping
of all Majoranas leaves even the SYK Hamiltonian invariant.  Thus, the
b-SYK Hamiltonian has a single extra  $\Ztwo$ symmetry compared to the
SYK Hamiltonian.  This explains our observation of a systematic shift
of level statistics, described in the previous subsection.

The operator $\Gamma$ can also be regarded as a particle-hole
conjugation operator.  If each $A$ Majorana is paired with a $B$
Majorana so that the Hilbert space is expressed in terms of complex
(usual) fermion Fock space, as described in Appendix
\ref{app_numerics}, then flipping signs of $a^A_j$'s amounts to a
transmutation of creation operators of complex fermions into
annihilation operators and vice versa, i.e., a particle-hole
conjugation.  (In terms of the original Majoranas, the $\Gamma$
operator changes a $n$-Majorana state $a^A_1 a^A_2\ldots a^A_n\ket{0}$
to a $(N_A-n)$-Majorana state
$a^A_{n+1} a^A_{n+2}\ldots a^A_{N_A}\ket{0}$.)
Thus, the $\Ztwo$ symmetry can be regarded as a particle-hole
conjugation symmetry, if we use the representation that each $A$
Majorana is paired with a $B$ Majorana.

We note parenthetically that the b-SYK Hamiltonian also admits a
number of operations that leave the Hamiltonian isospectral, although
not invariant: Exchanging any one of the $A$-fermions with any one of
the $B$-fermions leaves the Hamiltonian isospectral, i.e., amounts to
unitary operations.  This emerges due to the restriction from SYK to
b-SYK: For the SYK Hamiltonian, such operations are not isospectral.
For both SYK and b-SYK, bi-partitioning the Majorana fermions into
arbitrary halves and then exchanging the two halves is an isospectral
(unitary) operation.  The extra feature of the b-SYK is that
exchanging a single $A$ fermion with a single $B$ fermion is also a
unitary operation.

\begin{figure*}[t]
\begin{centering}
\includegraphics[width=0.95\linewidth]{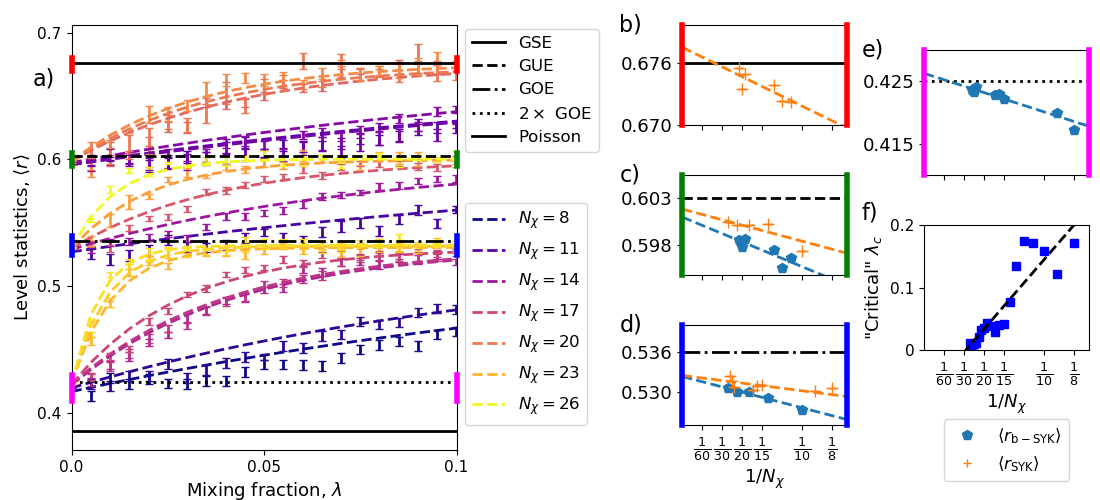}
\par\end{centering}
\caption{Level statistics when tuning between $\HBSYK$ and $\HSYK$   Hamiltonians,
  with  $\lambda=0$ ($\lambda=1$) corresponding to b-SYK (SYK).    
  In panel a) the average spacing ratio $\langle{r}\rangle$ is plotted as a function of $\lambda$.
  The crossover from  $\rbSYK$ to $\rSYK$ is steeper for a larger number of Majoranas $N_\chi$.
  Panels b) - e) zoom in on the regions around the various values of $\rbSYK$ and $\rSYK$ that are obtained.
  The colors on the vertical axes should help to match with the corresponding regions in panel a).
  These plots b) - e) show that the large-size limits are consistent with the GSE, GUE, GOE, and $2\times$GOE classes.     
  In panel a) a fit is performed to the function $\rSYK + (\rbSYK -\rSYK )  e^{-\frac{\lambda}{\lambda_C}}$ and on panel f) the transition parameter $\lambda_C$ is shown.
  The ``critical'' $\lambda_C$ decreases with increasing $N_\chi$. 
  \label{fig:Level_Stat_hybrid}}
\end{figure*}

\subsection{Interpolation between $\HBSYK$ to $\HSYK$}\label{sec:interpol}

Since the level statistics classification is systematically shifted from $\HSYK$ to $\HBSYK$ this begs the question what one obtains for a mixture of the two Hamiltonians.
We therefore define an interpolation Hamiltonian
\[
H_{\text{Mix}}=\left(1-\lambda\right)\HBSYK+\lambda\HSYK
\]
and investigate its level statistics as a function of $\lambda$.
In the following analysis, we choose the coupling constants $J$ in \eqref{eq:SYK} and \eqref{eq:bSYK} such that the variance of $J_{ijlm}$ in both cases is unity for all system sizes.
The main results are summarized in Fig.~\ref{fig:Level_Stat_hybrid}.
In panel a) we demonstrate how for already a small SYK-mixing, $\lambda \leq 0.1$, results in a drift of the level statistics from the $\HBSYK$ to the $\HSYK$ random matrix class.
This makes sense because, as soon as interactions are allowed which violate the bipartite restriction, the additional symmetry of the b-SYK Hamiltonian is lost.
The crossover happens faster (at even smaller values of $\lambda$) for larger system sizes, indicating that, in the large-$N_{\chi}$ limit,
an infinitesimal influence of $\HSYK$ is enough to move the system into the lower-symmetry class of the un-restricted SYK Hamiltonian.

To quantify this size dependence of the  $\HBSYK\to\HSYK$ crossover, we fit $\rEXP$ to the function
\[
f(\lambda) = \rSYK + (\rbSYK -\rSYK ) e^{-\frac{\lambda}{\lambda_C}}
\]
such that $f(0)=\rbSYK$, and $f(\lambda\gg \lambda_C)\to \rSYK$.
$\lambda_C$ is the value for which $f(\lambda_C)=\rSYK$ to first order, \ie $f(0)+\lambda_C f^{\prime}(0) = \rSYK$.
This function captures the transition from $\rbSYK$ to $\rSYK$ as a function of $\lambda$.
The best fit is shown in dashed lines, and the numerical $\rEXP$ is shown with statistical errors.

The shift from $\rbSYK$ to $\rSYK$ takes place at smaller values of $\lambda$ if more Majorana fermions, $N_\chi$, are involved (brighter colors).
In panel f), this is further illustrated: we plot $\lambda_C$ as a function of inverse system size $1/N_\chi$.
Clearly, $\lambda_C$ tends to zero in the large $N_\chi$ limit, quantifying the intuition that the shift of behavior happens at smaller $\lambda$ for larger sizes.

In panels b)-e) we show a scaling analysis for $\rbSYK$ and $\rSYK$, using the values at $\lambda=0$ and $\lambda=1$.
For $\rSYK$ (orange plus symbols), the large-size limit is consistent with the
known symmetry classes, GOE, GUE, or GSE, based on the value of $N_{\chi}$ mod 8 \cite{You2017,
  Garcia_Verbaarschot_PRD2017, Cotler_Polchinski_Shenker_Stanford_Tezuka_JHEP2017, Haque2019}.  For
$\rbSYK$ (blue circles), the large-size limit is consistent with the values corresponding to
$2\times$GOE, GOE, or GUE.
In panel b), focusing on the GSE value $\langle{r}\rangle\approx0.67$,
only $\lambda=1$ data (values of $\rSYK$) are visible.
Similarly, in panel e), focusing on the $2\times$GOE value $\langle{r}\rangle\approx0.425$,
only $\lambda=0$ data (values of $\rbSYK$) are visible.

\section{Discussion \& Context}\label{sec:results}

This paper has studied a bipartite version of the quartic ($q=4$) SYK model, which we call b-SYK.
It consists of two flavors of Majorana fermions that interact between the sets, but not within --- each quartic interaction term involves two Majorana fermions from one set and two from the other set.
The model was motivated in Ref.~\cite{Fremling2021} as being realizable in a specific setup of a strained version of the Kitaev honeycomb model.

Variants of the SYK model with two species of fermions have appeared previously, perhaps most prominently with the motivation of modeling eternal traversable wormholes using two quartic SYK models with only quadratic interactions between them
\cite{Milekhin_2021, Maldacena_Qi_arxiv2018_eternal,
  GarciaGarcia_Nosaka_Rosa_Verbaarschot_PRD2019, Plugge_Franz_PRL2020_revival,
  Sahoo_Franz_PRR2020_wormhole_coupledcomplex, Nosaka_Numasawa_JHEP2021,
  Franz_PRD2021_TraversableWormhole, Alet_Hanada_JHEP2021, Maldacena_Milekhin_JHEP2021,
  GarciaGarcia_Zheng_Ziogas_PRD2021, PengfeiZhang_JHEP2021_complex_SYKwormholes}.
In Ref.~\cite{Kim2019} the coupling between the two SYK clusters is quartic like ours.
Since our b-SYK model has no internal coupling within the two sets, it may be regarded as an infinite-coupling limit of the model of Ref.~\cite{Kim2019}, i.e., the limit in which the intra-set couplings can be neglected.
Ref.~\cite{Klebanov_Milekhin_Tarnopolsky_Zhao_JHEP2020} treats a complex-fermion version.
Ref.~\cite{PengfeiZhang_JHEP2017_sykbath} also considers two SYK clusters and quartic couplings between them, but the sizes of the two clusters are parametrically different, so that one acts as a bath for the other.
Several other two-flavor or two-species SYK variants have also appeared in the literature
\cite{Banerjee_Altman_PRB2017, Haldar_Shenoy_PRB2018,
  Haldar_etal_PRR2020_quench, Haldar_Tavakol_Scaffidi_PRR2021}.

We study the b-SYK model both analytically and numerically.
We find that in the large-$N$ limit, the model remains asymptotically solvable, showing conformal invariance in the infrared.
We establish that if we keep the ratio between the flavors a variable, we can continuously tune the scaling dimension of the respective species between $0$ and $1/2$.

For finite system sizes, we analyze the level statistics of the model numerically for
$\kappa=1$ ($N_A=N_B$ or $N_A=N_B\pm1$) and compare it to the known level statistics of the SYK model.
We find that the level statistics deviates systematically, consistent with the b-SYK model possessing an additional $\Ztwo$ symmetry.
The GOE, GUE, and GSE level statistics of the SYK model are reduced to $2\times$GOE, GOE, and GUE classes.

Studying the interpolation between the two models, we find that, for finite sizes,
the statistics evolve smoothly from the b-SYK to the SYK as a function of interpolating parameter $\lambda$.  

In the quantum chaos literature and in random matrix theory, the GOE-GUE crossover has been studied
repeatedly in various contexts~\cite{Pandey_Mehta_CommMathPhys1983,
  French_Kota_Pandey_Tomsovic_AnnalsPhys1988, Lenz_Haake_PRL1990, Lenz_Haake_PRL1991,
  Shukla_Pandey_Nonlinearity1997, Guhr_MuellerGroeling_WeidenMueller_PhysRep1998,
  Chung_etal_AnlageGroup_PRL2000, Haake_book2010, Schierenberg_Bruckmann_Wettig_PRE2012,
  Schweiner_Main_Wunner_PRE2017_GOEGUEPoissonTransitions,
  Schweiner_Laturner_Main_Wunner_PRE2017_crossover, Sarkar_Kothiyal_Kumar_PRE2020_GOEGUEcrossover,
  Corps_Relano_PRE2020}.
In the present case, we have a GUE to GSE crossover, a GOE to GUE crossover, and a $2\times$GOE to GOE crossover, all in the same Hamiltonian, depending on the number of Majorana fermions, according to the Bott periodicity
\cite{You2017, Garcia_Verbaarschot_PRD2017, Cotler_Polchinski_Shenker_Stanford_Tezuka_JHEP2017, Haque2019}.
In addition, unlike typical models studied in traditional quantum chaos or random matrix theory, we have a well-defined thermodynamic (large $N_{\chi}$) limit.
It turns out that, in this limit, the crossover happens extremely rapidly, i.e., the b-SYK statistics is lost already for an infinitesimal mixture of SYK.

The present work opens up a number of new questions.
Thermodynamic and thermalization properties, as well as higher-order correlation functions, and Lyapunov exponents, remain to be studied.
It may be interesting to see how b-SYK physics is explicitly obtained in the large-interaction limit of the model of Ref.~\cite{Kim2019},
and to investigate the behavior of its complex-fermion version.
In addition, the level statistics for unequal-sized bipartitions ($\kappa\neq1$) also deserves exploration.

\section*{Acknowledgements}

We would like to thank Philippe Corboz, Maria Hermanns, Lukas Janssen, Graham Kells, Tobias Meng,
Subir Sachdev, Alexey Milekhin, and Matthias Vojta for useful discussions.
This work is part of the D-ITP
consortium, a program of the Netherlands Organisation for Scientific Research (NWO) that is funded
by the Dutch Ministry of Education, Culture and Science (OCW).

\appendix

\section{The scaling dimensions}\label{app_scaling_dims}

In this Appendix, we show how equations \eqref{eq:Dysoneqs_B} can be
used to derive the relationship, Eq.\ \eqref{eq_scalin_dims}, between the ratio
$\kappa= N_A/N_B$ and (one of) the scaling dimensions. 

Dividing one of equations \eqref{eq:Dysoneqs_B} by the other gets rid
of $\Lambda$.  We also use $\Delta_{B}=\frac{1}{2}-\Delta_{A}$, to obtain
\begin{equation} \label{eq:Dysoneqs_C}
\kappa =  \frac{\Gamma\left(\Delta_{A}\right)\Gamma\left(1+\Delta_{A}\right)}{\Gamma\left(\frac{1}{2}-\Delta_{A}\right)\Gamma\left(\frac{3}{2}-\Delta_{A}\right)}\frac{\Gamma^{2}\left(1-\Delta_{A}\right)}{\Gamma^{2}\left(\frac{1}{2}+\Delta_{A}\right)}.
\end{equation}
Using Euler's reflection formula
\begin{equation}
\Gamma\left(1-z\right)\Gamma\left(z\right)=\frac{\pi}{\sin\left(\pi z\right)},\label{eq:F_1_z_F_z_identity}
\end{equation}
we find that 
\begin{align}
  \Gamma\left(1-\Delta_{A}\right)\Gamma\left(\Delta_{A}\right)
&= \frac{\pi}{\sin\left(\pi\Delta_{A}\right)}, \label{eq:F_F_to_sin}\\
 \Gamma\left(\frac{1}{2}-\Delta_{A}\right)\Gamma\left(\frac{1}{2}+\Delta_{A}\right)
&= \frac{\pi}{\cos\left(\pi\Delta_{A}\right)}. \label{eq:F_F_to_cos}
\end{align}
These can be used for pairs of products of Gamma functions in Eq.\
\eqref{eq:Dysoneqs_C}, yielding 
\begin{equation}
\kappa=  \frac{\Gamma\left(1+\Delta_{A}\right)}{\Gamma\left(\frac{3}{2}-\Delta_{A}\right)}\frac{\Gamma\left(1-\Delta_{A}\right)}{\Gamma\left(\frac{1}{2}+\Delta_{A}\right)}\frac{\cos\left(\pi\Delta_{A}\right)}{\sin\left(\pi\Delta_{A}\right)}.
\end{equation}
There remains pairs of Gamma functions in the denominator and numerator,
which we now proceed  to eliminate. 
By making use of the recursive property $\Gamma\left(1+z\right)=z\Gamma\left(z\right)$ we can write 
\begin{align*}
\Gamma\left(1+\Delta_{A}\right) & =\Delta_{A}\Gamma\left(\Delta_{A}\right), \\
\Gamma\left(\frac{3}{2}-\Delta_{A}\right) & 
= \left(\frac{1}{2}-\Delta_{A}\right)\Gamma\left(\frac{1}{2}-\Delta_{A}\right). 
\end{align*}
Applying these relations, and then again applying
\eqref{eq:F_F_to_cos} and \eqref{eq:F_F_to_sin}, leads to 
\begin{align*}
\kappa=
  \frac{2\Delta_{A}}{1-2\Delta_{A}}\frac{1}{\tan^{2}\left(\pi\Delta_{A}\right)}, 
\end{align*}
which is the desired result, Eq.\ \ref{eq_scalin_dims}.

\section{Numerical Calculations} \label{app_numerics}

In this Appendix we briefly describe the numerical procedure for
obtaining the b-SYK (and SYK) spectra, and make some technical remarks.

To form the basis for the Hilbert space, the $N_{\chi}$ Majorana
fermions are paired into complex or `usual' fermions (without spin).
For even $N_{\chi}$, this leads to $N_{\chi}/2$ complex fermions, and
hence the Hilbert space dimension is $2^{N_{\chi}/2}$.  This is a
manifestation of Majorana's representing half a fermionic degree of
freedom.  

Because the Hamiltonian is quartic, either in terms of the Majorana's
or in terms of the complex fermions, the odd-fermion states and the
even-fermion states fall into two disconnected sectors, which may be
diagonalized separately and have the same statistics.

One could explicitly introduce complex fermions.  For the SYK
Hamiltonian \eqref{eq:SYK}, we could pair the Majorana fermions as, e.g., 
\begin{equation}
\begin{split}
c_j^{\dagger} = \frac12 \left( \gamma_{j}+ \imath \gamma_{N_{\chi}-j+1} \right);
\\
c_j = \frac12 \left( \gamma_{j}- \imath \gamma_{N_{\chi}-j+1} \right). 
\end{split}
\end{equation}
One can then construct a basis for the Hamiltonian as the Fock basis of
these complex fermions, including the vacuum, the one-particle states,
the two-particle states, etc.:
\begin{equation}
\begin{split}
\ket 0, \\
c_1^{\dagger}\ket 0, c_2^{\dagger}\ket 0, \ldots, \\
c_1^{\dagger}c_2^{\dagger}\ket 0, c_1^{\dagger}c_3^{\dagger}\ket 0, \ldots, \\
\vdots \\
c_1^{\dagger}c_2^{\dagger}\ldots c_{N_{\chi}/2}^{\dagger} \ket 0 . 
\end{split}
\end{equation}
The Hamiltonian (with a particular realization of the random couplings
$J_{ijlm}$) can then be represented as a matrix in this basis.  It is
efficient to construct the matrices separately for the
even-occupancy and odd-occupancy sectors, since they are decoupled.
Either or both of these $2^{N_{\chi}/2-1}\times 2^{N_{\chi}/2-1}$ matrices
can then be numerically diagonalized to obtain the spectrum.  To
obtain sufficient statistics, this procedure is repeated for many
different realizations of the random couplings, and the data for level
spacings (or level spacing ratios) are aggregated. 

In practice, it is not necessary to explicitly express the Hamiltonian
in terms of complex fermions.  Combining $ \gamma_{j}$ and
$\gamma_{N_{\chi}-j+1}$ into a complex fermion mode is equivalent to
insisting that the vacuum has the property that $\gamma_{j}\ket 0$ and
$\imath\gamma_{N_{\chi}-j+1}\ket 0$ are the same state, for each
$j=1,2,\ldots,\frac{N_{\chi}}{2}$.  Thus, one can express the basis
states as
\begin{multline}
\ket 0, \;
\gamma_1\ket 0, \gamma_2\ket 0, \ldots, \;
\gamma_1\gamma_2\ket 0, \gamma_1\gamma_3\ket 0, \ldots, \\
\ldots , \quad 
\gamma_1\gamma_2\ldots \gamma_{N_{\chi}/2} \ket 0 . 
\end{multline}
and the Hamiltonian matrix elements is calculated directly in this
basis. 

For the b-SYK Hamiltonian, the procedure is exactly the same.  The
choice of how the $N_{\chi}$ Majorana fermions are divided into pairs should not
matter.  We choose to pair each $A$ Majorana with a $B$ Majorana:
\begin{equation}
\begin{split}
c_j^{\dagger} = \frac12 \left( a^B_{j} - \imath a^A_j\right);
\\
c_j = \frac12 \left( a^B_{j} + \imath a^A_{j}  \right). 
\end{split}
\end{equation}
This is equivalent to imposing on the vacuum $\ket 0$ the property
that $a^A_{j}\ket 0 = \imath a^B_{j}\ket 0$.

The symmetry operation discussed in subection \ref{sec:which_Z2},
flipping signs of all the $A$ Majorana's, corresponds to the
transformation $c_j^{\dagger}\leftrightarrow c_j$, i.e., a
particle-hole conjugation, in this representation.   

When $N_{\chi}$ is odd, there is a single unpaired Majorana fermion.
In this case, one simply adds a fictitious additional Majorana to form
the last pair.  For example, in the b-SYK case, if $N_B=N_A-1$, we add
the fictitious Majorana operator $a^B_{N_B+1}=a^B_{N_A}$, which never appears in
the Hamiltonian, and impose $a^A_{N_A}\ket 0 = \imath a^B_{N_A}\ket
0$.  

A final remark: For the symplectic cases, the spectrum has a two-fold
degeneracy for both the SYK and b-SYK hamiltonians.  Therefore half of
the energy levels must be pruned to get rid of this ``trivial''
symmetry from the spectrum.

\bibliography{KitaevShortBib}

\end{document}